\pdfoutput=1

\documentclass[11pt,twoside,a4paper,cmspaper,final,collab]{cms-tdr}
\def\svnVersion{435508P}\def\cmsCernNoTag{CERN-EP-2017-097}\def\cmsCernDate{\today}\def\cmsMessage{Published in the Journal of High Energy Physics as \href{http://dx.doi.org/10.1007/JHEP10(2017)073}{\doi{10.1007/JHEP10(2017)073.}}}

\begin{document}\cmsNoteHeader{EXO-16-039}

\hyphenation{had-ron-i-za-tion}
\hyphenation{cal-or-i-me-ter}
\hyphenation{de-vices}
\RCS$Revision: 420448 $
\RCS$HeadURL: svn+ssh://svn.cern.ch/reps/tdr2/papers/EXO-16-039/trunk/EXO-16-039.tex $
\RCS$Id: EXO-16-039.tex 420448 2017-08-10 09:30:30Z yiiyama $
\newlength\cmsFigWidth
\ifthenelse{\boolean{cms@external}}{\setlength\cmsFigWidth{0.85\columnwidth}}{\setlength\cmsFigWidth{0.4\textwidth}}
\ifthenelse{\boolean{cms@external}}{\providecommand{\cmsLeft}{top\xspace}}{\providecommand{\cmsLeft}{left\xspace}}
\ifthenelse{\boolean{cms@external}}{\providecommand{\cmsRight}{bottom\xspace}}{\providecommand{\cmsRight}{right\xspace}}

\newcommand{\currentL}{12.9}
\newcommand{\currentLuncert}{0.8}
\newcommand{\mew}{\ensuremath{M_{\mathrm{EW}}}}
\newcommand{\mpl}{\ensuremath{M_{\mathrm{Pl}}}}
\newcommand{\mD}{\ensuremath{M_D}}
\newcommand{\mdm}{\ensuremath{m_{\text{DM}}}}
\newcommand{\mmed}{\ensuremath{M_{\text{med}}}}
\newcommand{\gDM}{\ensuremath{g_{\text{DM}}}}
\newcommand{\gq}{\ensuremath{g_{\cPq}}}
\newcommand{\ns}{\ensuremath{\,\text{ns}}\xspace}
\newcommand{\met}{\ptmiss}
\newcommand{\MT}{\ensuremath{M_{\mathrm{T}}}}
\newcommand{\dR}{\ensuremath{\Delta R}}
\newcommand{\gmet}{\ensuremath{\gamma + \met}}
\newcommand{\zinvg}{\ensuremath{\PZ(\to\PGn\PAGn)+\Pgg}}
\newcommand{\zllg}{\ensuremath{\PZ(\to\ell\overline{\ell})+\Pgg}}
\newcommand{\zllj}{\ensuremath{\PZ(\to\ell\overline{\ell})+\text{jets}}}
\newcommand{\wlng}{\ensuremath{\PW(\to\ell\PGn)+\Pgg}}
\newcommand{\gj}{\ensuremath{\Pgg+\text{jets}}}
\newcommand{\ttg}{\mbox{\ttbar\hspace{-0.3em}\Pgg}\xspace}
\newcommand{\ETg}{\ensuremath{p_{\mathrm{T}}^{\Pgg}}}
\newcommand{\sigetaeta}{\ensuremath{\sigma_{\eta\eta}}}
\newcommand{\as}{\ensuremath{\alpha_s}}
\newcommand{\muR}{\ensuremath{\mu_{R}}}
\newcommand{\muF}{\ensuremath{\mu_{F}}}
\newcommand{\DphiMETg}{\ensuremath{\Delta \phi(\MET, \Pgg)}}
\newcommand{\DphiMETj}{\ensuremath{\Delta \phi(\ptvecmiss, {\vec p}_{\mathrm{T}}^{\kern1pt\text{jet}})}}
\newcommand{\minDphiMETj}{\ensuremath{\mathrm{min}\DphiMETj}}
\newcommand{\MGfiveAMC} {\MADGRAPH{}5\_\allowbreak a\MCATNLO}
\newcommand{\NNPDFthree}{{NNPDF 3.0}\xspace}
\newcommand{\DYRes}{\textsc{DYRes}\xspace}

\newcommand{\x}{\ensuremath{\phantom{0}}}

\hyphenation{NN-PDF}
\hyphenation{PY-THIA}

\cmsNoteHeader{EXO-16-039}
\title{Search for new physics in the monophoton final state in proton-proton collisions at $\sqrt{s}=13\TeV$}

\date{\today}

\abstract{
A search is conducted for new physics in a final state containing a photon and missing transverse momentum in proton-proton collisions at $ \sqrt{s} = 13\TeV $. The data collected by the CMS experiment at the CERN LHC correspond to an integrated luminosity of 12.9\fbinv. No deviations are observed relative to the predictions of the standard model. The results are interpreted as exclusion limits on the dark matter production cross sections and parameters in models containing extra spatial dimensions. Improved limits are set with respect to previous searches using the monophoton final state. In particular, the limits on the extra dimension model parameters are the most stringent to date in this channel.
}

\hypersetup{%
pdfauthor={CMS Collaboration},%
pdftitle={Search for dark matter and graviton produced in association with a photon in pp collisions at sqrts = 13 TeV},%
pdfsubject={CMS},%
pdfkeywords={CMS, physics, software, computing}}

\maketitle

\section{Introduction}

One of the most intriguing open questions in physics is the nature of dark matter (DM). While DM is
thought to be the dominant nonbaryonic contribution to the matter density of the
universe~\cite{Gaitskell:2004gd}, its detection and identification in terrestrial and spaceborne
experiments remains elusive. At the CERN LHC, the DM particles may be produced in high-energy
proton-proton collisions, if the DM particles interact with the standard model (SM) quarks or gluons
via new couplings at the electroweak scale~\cite{DMLHC1,DMCollid}. Although DM particles cannot be
directly detected at the LHC, their production could be inferred from an observation of events with
a large transverse momentum imbalance (missing transverse momentum, \ptmiss, defined in
Section~\ref{sec:detector}).

Another highly important issue is the hierarchy problem, which involves the large energy gap between
the electroweak (\mew) and Planck (\mpl) scales~\cite{EWbreak}. Proposed solutions to this problem
include theories with large extra dimensions, such as the model of Arkani-Hamed, Dimopoulos, Dvali
(ADD)~\cite{ADD,ADD1}. The ADD model postulates that there exist $n$ compactified extra dimensions
in which gravitons can propagate freely and that the true scale (\mD) of the gravitational
interaction in this $4{+}n$ dimensional space-time is of the same order as \mew. The compactification
scale $R$ of the additional dimensions is related to the two gravitational scales by $\mpl^{2} \sim
R^{n}\mD^{n+2}$. For $\mD \sim \mew$, the cases $n=1$ and $n=2$ are ruled out or strongly disfavored by
various observations~\cite{ADD1}, while cases $n \geq 3$ remain to be probed, for example, by collider experiments.
The compactification scale $R$ is much greater than $1/\mew$ for a wide range of $n$, leading to a
near-continuous mass spectrum of Kaluza--Klein graviton states. Although the gravitons would not be
observed directly at the LHC, their production would be manifest as events broadly distributed in
\ptmiss.

\begin{figure}[hbtp]
\centering
\includegraphics[width=0.28\linewidth]{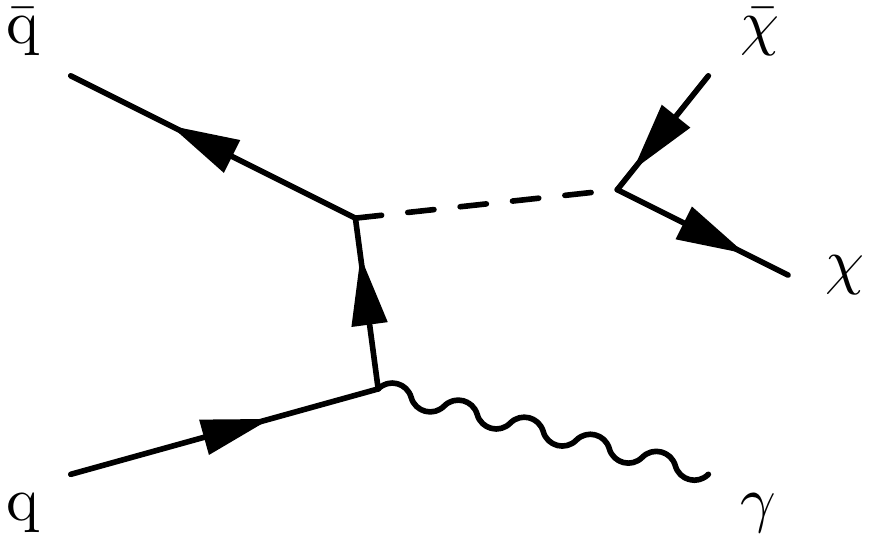}
\includegraphics[width=0.28\linewidth]{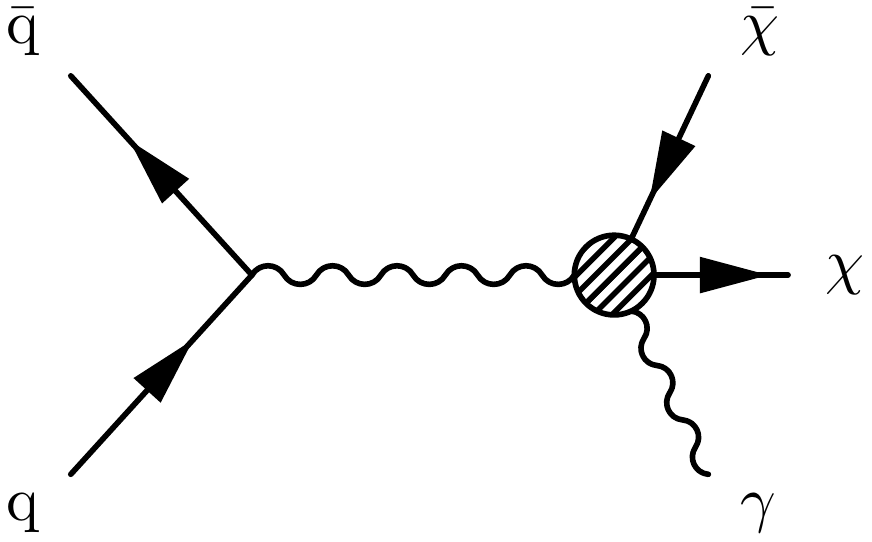}
\includegraphics[width=0.28\linewidth]{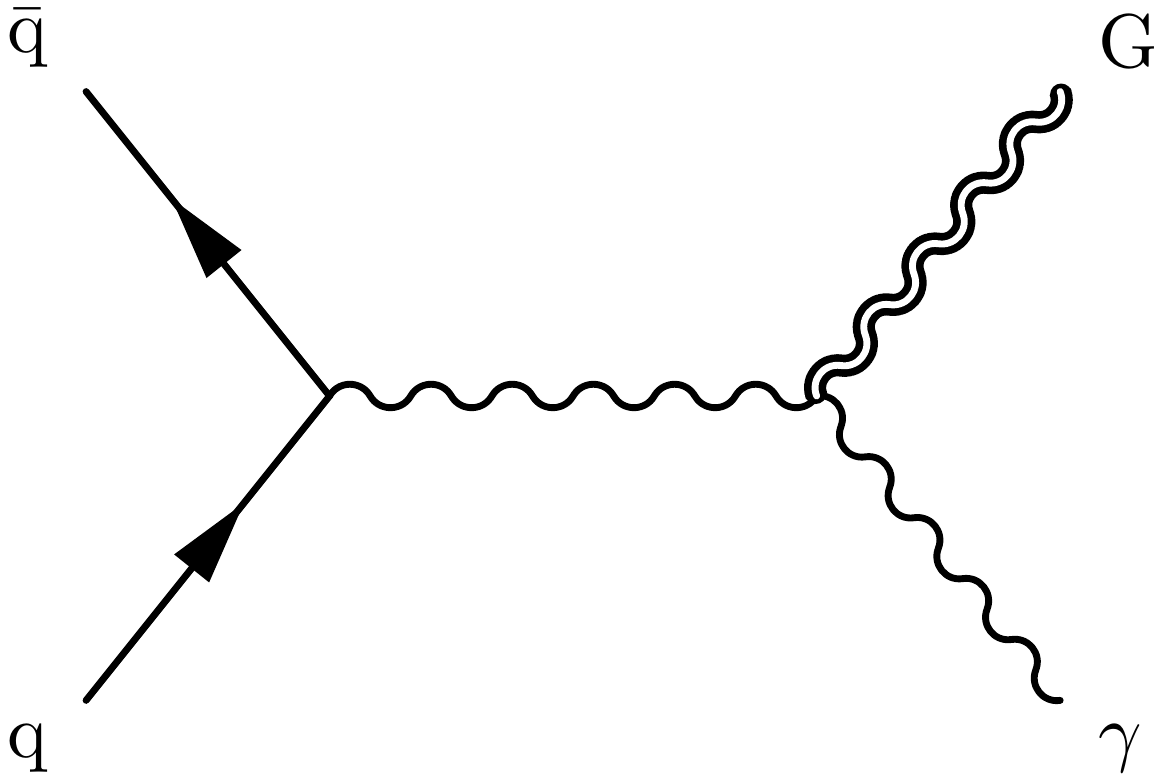}
\caption{
Leading-order diagrams of the simplified DM model (left), electroweak-DM effective interaction (center), and graviton (G) production in the ADD model (right), with a final state of \Pgg\ and large \ptmiss.
}
\label{fig:diagrams}

\end{figure}

In generic models of DM and graviton production, various SM particles can recoil against these
undetected particles, producing a variety of final states with significant \ptmiss. The monophoton,
or $\Pgg+\ptmiss$, final state has the advantage of being identifiable with high efficiency and
purity. In DM production through a vector or axial vector mediator, a photon can be radiated from
incident quarks (Fig.~\ref{fig:diagrams} left). Models of this process have been developed by the
CMS-ATLAS Dark Matter Forum~\cite{dmforum}. It is also possible that the DM sector couples
preferentially to the electroweak sector, leading to an effective interaction
$\PQq\PAQq\to\PZ/\Pgg^*\to\Pgg\chi\overline{\chi}$~\cite{Nelson:2013pqa}, where $\chi$ is the DM particle
(Fig.~\ref{fig:diagrams} center). In ADD graviton production, the graviton can couple directly to
the photon (Fig.~\ref{fig:diagrams} right) or to a quark. In this paper, we examine final states
containing large \ptmiss in the presence of a photon with large transverse momentum (\PT), and
search for an excess of events over the SM prediction. Data collected by the CMS experiment in 2016
with an integrated luminosity of \currentL\fbinv are analyzed. Results are interpreted in the
context of these three models.

The primary irreducible background for the $\Pgg+\ptmiss$ signal is SM \PZ boson production associated with a photon, \zinvg. Other SM backgrounds include \wlng\ (having a final state photon, and a lepton $\ell$ that escapes detection), $\PW\to\ell\Pgn$ (where $\ell$ is misidentified as a photon), \gj, quantum chromodynamics (QCD) multijet events (with a jet misidentified as a photon), \ttg, VV\Pgg\ (where V refers to a \PW\ or a \PZ boson), \zllg, and noncollision sources, such as beam halo interactions and detector noise.

A previous search in the $\Pgg+\ptmiss$ final state using \Pp\Pp\ collisions at $\sqrt{s}=8\TeV$, corresponding to an integrated
luminosity of 19.6\fbinv, was reported by the CMS experiment in Ref.~\cite{mono8TeV}. The ATLAS experiment has also reported a similar search in 3.2\fbinv of \Pp\Pp\ collisions at $\sqrt{s} = 13\TeV$~\cite{Aaboud:2016uro}.

\section{The CMS detector and candidate reconstruction}
\label{sec:detector}

The central feature of the CMS apparatus is a superconducting solenoid of 6\unit{m} internal
diameter, providing a magnetic field of 3.8\unit{T}. Within the solenoid volume are
a silicon pixel and strip tracker, a lead tungstate crystal electromagnetic calorimeter (ECAL), and
a brass and scintillator hadron calorimeter (HCAL), each composed of a barrel ($\abs{\eta}< 1.48$)
and two endcap ($1.48 <\abs{ \eta } < 3.00$) sections, where $\eta$ is the pseudorapidity. Extensive
forward calorimetry complements the coverage provided by the barrel and endcap detectors. Muons are
measured in gas-ionization detectors embedded in the steel flux-return yoke outside the solenoid.

An energy resolution of about 1\% is reached within the barrel section of the ECAL for unconverted or
late-converting photons with $\PT \geq 60\GeV$. The remaining barrel photons have a
resolution of about 1.3\% up to a pseudorapidity of $|\eta| = 1$, rising to about 2.5\% at $|\eta| =
1.4$~\cite{Khachatryan:2015iwa}. The time resolution of photons at the ECAL is ${<}200 \unit{ps}$ for
depositions ${>} 10\GeV$. In the $\eta$--$\phi$ plane, where $\phi$ is the azimuthal angle,
and for $\abs{\eta}< 1.48$, the HCAL cells map onto $5 {\times} 5$ arrays of ECAL crystals to form
calorimeter towers projecting radially outward from the center of the detector. A more detailed
description of the CMS detector, together with a definition of the coordinate system and kinematic variables, can be found in Ref.~\cite{Chatrchyan:2008aa}.

Events of interest are selected using a two-tiered trigger system~\cite{Khachatryan:2016bia}. The
first level (L1), composed of custom hardware processors, uses information from the calorimeters and
muon detectors to select events at a rate of around 100\unit{kHz} within a time interval of less
than 4\mus. The second level, known as the high-level trigger (HLT), consists of a farm of
processors running a version of the full event reconstruction software optimized for fast
processing, and reduces the event rate to less than 1\unit{kHz} before data storage.

Event reconstruction is performed using a particle-flow (PF)
technique~\cite{CMS-PAS-PFT-09-001,CMS-PAS-PFT-10-002}, which reconstructs and identifies individual
particles using an optimized combination of information from all subdetectors. The energy of charged
hadrons is determined from a combination of the track momentum and the corresponding ECAL and HCAL
energies, corrected for the combined response function of the calorimeters. The energy of neutral
hadrons is obtained from the corresponding corrected ECAL and HCAL energies. Muon identification and
momentum measurements are performed by combining the information from the inner trackers and outer muon
chambers.

The PF candidates in each event are clustered into jets via the anti-$k_{t}$
algorithm~\cite{Cacciari:2008gp,Cacciari:2011ma} with a distance parameter of 0.4. Jet energies, computed from a
simple sum of 4-momenta of the constituent PF candidates, are corrected to
account for the contributions from particles associated with additional interactions within the same or nearby
bunch crossings (pileup), as well as to compensate for the nonlinearities in the measured particle
energies. Jet energy corrections are obtained from simulation, and are confirmed through in situ
measurements of the energy momentum balance in dijet and photon + jet events.

The uncorrected missing transverse momentum vector (\ptvecmiss) is
defined as the negative vector sum of the transverse momenta of all PF
candidates in an event. This quantity is adjusted with the difference
of uncorrected and corrected jets for a consistent and more accurate
missing momentum measurement~\cite{Khachatryan:2014gga}. The magnitude
of \ptvecmiss\ is referred to as the missing transverse
momentum, \ptmiss.

The reconstruction of photons and electrons begins with the identification of clusters of energy
deposited in the ECAL with little or no observed energy in the corresponding HCAL region. For each
candidate cluster, the reconstruction algorithm searches for hits in the pixel and strip trackers that can be associated with the cluster. Such associated hits are called electron seeds, and are used to
initiate a special track reconstruction based on a Gaussian sum
filter~\cite{Adam:2003kg,Khachatryan:2015hwa}, which is optimized for electron tracks.
A ``seed veto'' removes photon candidates with an associated electron seed.

Selections based on calorimetric information and isolation are applied to distinguish photons
from electromagnetic (EM) showers caused by hadrons. The calorimetric requirements for photons comprise $H/E <
0.05$ and $\sigetaeta < 0.0102$, where $H/E$ is the ratio of hadronic to EM energy
deposition. The variable $\sigetaeta$, described in detail in Ref.~\cite{Khachatryan:2015iwa},
represents the width of the electromagnetic shower in the $\eta$ direction, which is generally
larger in showers from hadronic activity. For a photon candidate to be considered as isolated, scalar sums
of the transverse momenta of PF charged hadrons, neutral hadrons, and photons within a cone of $\dR
= \sqrt{\smash[b]{(\Delta \eta)^2 + (\Delta \phi)^2}} < 0.3$ around the candidate photon must individually fall
below the bounds defined for 80\% signal efficiency. Only the PF candidates that do not overlap with the
EM shower of the candidate photon are included in the isolation sums.

Each PF charged hadron is reconstructed from a track and can be associated with an interaction
vertex it originates from. Therefore, the isolation sum over PF charged hadrons should be computed
using only the candidates sharing an interaction vertex with the photon candidate. However, because
photon candidates are not reconstructed from tracks, their vertex association is ambiguous. When an
incorrect vertex is assigned, photon candidates that are not isolated can appear otherwise. To
mitigate the rate for accepting nonisolated candidates as photon candidates, the maximum charged hadron
isolation value over all vertex hypotheses (worst isolation) is used.

Another consequence of calorimetry-driven reconstruction is that stray ECAL clusters produced by mechanisms other than
\Pp\Pp\ collisions can be misidentified as photons. In particular, beam halo muons that accompany proton beams and penetrate the detector longitudinally, and
the interaction of particles in the ECAL photodetectors (``ECAL spikes'') have been found to
produce spurious photon candidates at nonnegligible rates. To reject these backgrounds, the ECAL
signal in the seed crystal of the photon cluster is required to be within ${\pm} 3\ns$ of the arrival time
expected for particles originating from a collision. In addition, the candidate cluster must comprise
more than a single ECAL crystal. Furthermore, the maximum of the total
energy along all possible paths of beam halo particles passing through the cluster is
calculated for each photon candidate. This quantity, referred to as the halo total energy, is required to be below a threshold defined to
retain 95\% of the true photons, while rejecting 80\% of the potential halo clusters.

\section{Event selection}
\label{sec:event_selection}

The integrated luminosity of the analyzed data sample, derived from a preliminary measurement using the method described in~\cite{CMS-PAS-LUM-16-001}, is
$(\currentL\pm\currentLuncert)\fbinv$. The data sample is collected with a single-photon trigger
that requires at least one photon candidate with $\pt >165\GeV$. The photon candidate must have $H/E
< 0.1$, to reject jets. The photon energy reconstructed in the trigger is less precise relative to
that derived later in the offline selection. Therefore, the thresholds in the trigger on both $H/E$ and \ETg, where
\ETg\ is the photon \pt, are less restrictive than their offline counterparts.  The trigger efficiency is
measured to be about 98\% for events passing the analysis selection with $\ETg >175\GeV$.

From the recorded data, events are selected by requiring $\ptmiss > 170\GeV$ and at least one photon
with $\ETg > 175\GeV$ in the fiducial region of the ECAL barrel ($\abs{\eta} < 1.44$). Events are
rejected if the minimum opening angle between \ptvecmiss and any of the four highest transverse
momenta jets, \DphiMETj, is less than 0.5. This requirement significantly suppresses spurious
\ptmiss backgrounds from mismeasured jets. Only jets with $\PT > 30\GeV$ and $|\eta| < 5$ are
considered in the \DphiMETj\ calculation. The candidate photon transverse momentum vector and
\ptvecmiss must be separated by more than 2 radians. Finally, to reduce the contribution from the
\wlng\ process, events are vetoed if they contain an electron or a muon with $\PT > 10\GeV$ that is
separated from the photon by $\dR > 0.5$.

\section{Signal and background modeling}
\label{sec:bkgestimate}

The SM backgrounds and signal are modeled using both simulated events and recorded data. The two methods are described in the following Sections.

\subsection{Monte Carlo simulation for signal and background modeling}

{\tolerance = 600
Monte Carlo (MC) simulation is used to model the signal and some classes of SM background events. For the SM backgrounds, the primary hard
interaction is simulated using the \MGfiveAMC\ version 2.2.2~\cite{Alwall:2014hca} or
\PYTHIA{}8.212~\cite{Sjostrand:2014zea} generators employing the \NNPDFthree~\cite{Ball:2014uwa} leading-order (LO) parton
distribution function (PDF) set at the strong coupling value $\alpha_{S} = 0.130$. Parton showering and hadronization are provided in
\PYTHIA{}8.212 through the underlying-event tune CUETP8M1~\cite{tune}. Multiple minimum-bias events are
overlaid on the primary interaction to model the distribution of pileup in data. Generated particles are processed through the
full \GEANTfour-based simulation of the CMS detector~\cite{GEANT, GEANTdev}.
\par
}

For the DM signal hypothesis, MC simulation samples are produced with \MGfiveAMC\ 2.2.2, requiring $\ETg > 130\GeV$ and
$\abs{\eta^{\Pgg}} <2.5$. A large number of DM simplified model samples are generated, varying the
masses of the mediator and DM particles. Similarly, electroweak-DM effective interaction samples are
generated with a range of dark matter masses. For the ADD hypothesis, events are generated using
\PYTHIA{}8.212, requiring $\ETg> 130\GeV$, with no restriction on the photon pseudorapidity. Samples
are prepared in a grid of number of extra dimensions and $M_{D}$. The efficiency of the full event
selection on these signal models ranges between 0.12 and 0.27 for the DM simplified models, 0.42 and
0.45 for electroweak DM production, and 0.22 and 0.28 for the ADD model, depending on the parameters
of the models.

Predictions for signal and background MC yields are rescaled by an overall correction factor
($\rho$) that accounts for the differences in event selection efficiency between data and
simulation. The value of $\rho= 0.94\pm0.06$ reflects the product of three correction factors:
$0.94\pm0.01$ for photon identification and isolation, $1.00\pm0.01$ for the electron seed veto, and
$1.00\pm0.06$ for the combination of the worst isolation, the BHM total energy requirement, and the lepton veto. The
selection efficiencies are measured in data using the tag-and-probe
technique~\cite{Khachatryan:2010xn}. Events with
$\PZ\to\Pe\Pe$ decays are employed for measuring the photon identification and isolation
efficiencies, while a $\PZ\to\Pgm\Pgm\Pgg$ sample is utilized to extract the other
efficiency factors~\cite{CMS:2011aa}.

The most significant SM backgrounds in this search are from the associated production of a \PZ\ or \PW\ boson
with a high-energy photon, denoted as \zinvg\ and \wlng. When the \PZ\ boson decays into a
neutrino-antineutrino pair, the final state exhibits a high-\PT photon and large \ptmiss. Similarly, if the \PW\ boson decays
into a lepton-neutrino pair and the lepton escapes detection, the event appears to be $\Pgg+\ptmiss$. Together, these processes account for
approximately 70\% of the SM background, with 50\% from \zinvg\ alone.

The estimation of \zinvg\ and \wlng\ backgrounds is based on
\MGfiveAMC\ simulations at LO in QCD and with up to two additional
partons in the final state. In addition to the selection efficiency
correction factor $\rho$, these samples are weighted event-by-event
with the product of two factors. The first factor matches the
distribution of the generator-level \ETg\ to that calculated at
next-to-next-to-leading order (NNLO) in QCD using the \DYRes
program~\cite{Catani:2015vma}. The second factor, taken from
Refs.~\cite{Denner:2014bna,Denner:2015fca}, further corrects the
backgrounds to account for next-to-leading order (NLO) electroweak effects. The estimated
contributions from the \zinvg\ and \wlng\ processes after applying the
selections in Section~\ref{sec:event_selection} are given in Table~\ref{tab:BkgSummaryC},
and amount to $215 \pm 32$ and $57.2 \pm 8.0$ events,
respectively. Statistical and systematic uncertainties are combined in
quadrature. The statistical uncertainty is subdominant and is due to
the finite size of the simulation sample. Systematic uncertainties in
the estimated \zinvg\ and \wlng\ yields have four contributions and are summarized in Table~\ref{tab:sysfull1}. The
first is associated with the PDF and the choice of renormalization and
factorization scales (\muR\ and \muF) used in generating the
events. The relative uncertainty from these sources are 5.4\% and
8.9\% in the \zinvg\ and \wlng\ yields, respectively. Uncertainty from
the PDF is evaluated by varying the weight of each event based on the
standard deviation of the event weight distribution as given by the
NNPDF set. Uncertainties from the choice of \muR\ and \muF\ are
evaluated by setting the scales to twice or half the nominal values
and taking the minima and maxima of the resulting event
weights. Second, the uncertainty due to missing higher-order
electroweak corrections is taken as the magnitude of the NLO
correction. The uncertainty from this source is 11\% for the
\zinvg\ process and 7\% for \wlng. The third uncertainty is on the
selection efficiency correction factor $\rho$, with the main contribution
from the statistical uncertainties in individual efficiency
measurements.  A fourth uncertainty is assigned to cover the
uncertainties in the jet energy scale~\cite{CMS:2017wyc}, photon
energy scale~\cite{CMS-PAS-EGM-10-006}, pileup, and the scale and
resolution in \met. The combined relative uncertainties from the third
and fourth categories in the \zinvg\ and \wlng\ yields are 6\% and 6.2\%,
respectively.

To validate the predictions from simulation, observed and MC simulated data are compared in two control regions.  One
region consists of events with two same-flavor leptons of opposite-charge and a photon, which is
dominated by the \zllg\ process. The photon is selected by criteria identical to those used in the signal
candidate event selection, while the leptons are required to have $\PT > 10\GeV$ and the dilepton
invariant mass must lie between 60 and 120\GeV. Furthermore, the recoil
$U^{\ell\overline{\ell}} = |\ptvecmiss + {\vec p}_{\mathrm{T}}^{\, \ell} +
{\vec p}_{\mathrm{T}}^{\, \overline{\ell}}|$~\cite{Khachatryan:2010xn} must be greater than 170\GeV to emulate the \met in \zinvg\ events. In addition to simulated
\zllg\ events, MC samples of \ttg, \zllj, and multiboson events are also considered. In total, $68.1
\pm 3.8$ events are predicted in the dilepton control region, and 64 events are observed. The
dominant uncertainty is theoretical. Using the ratio of acceptances between the \zinvg\ and
\zllg\ simulations, this validation is used to predict the \zinvg\ contribution to the
candidate sample of $242 \pm 35$, which is in agreement with the purely simulation-based prediction
given previously. The uncertainty in this prediction is mainly due to the limited event yields in
the control samples.

The second region is defined by requirements of exactly one electron
or muon with $\PT > 30\GeV$, one photon with $\PT > 175\GeV$, $\met >
50\GeV$, and $U^{\ell} = |\ptvecmiss + {\vec p}_{\mathrm{T}}^{\, \ell}| >
170\GeV$~\cite{Khachatryan:2014gga}. This region is dominated by
$\PW(\to\ell\PGn)+\Pgg$ production. A total of 108 events are observed
in this region, where $10.6 \pm 1.3$ non-$\PW+\Pgg$ background events
are expected. The ratio of the acceptance for $\PW+\Pgg$ events where
the lepton is missed, compared to the acceptance for events where it
is identified is estimated from simulation, and is multiplied with the background-subtracted observed
yield of this control region. The product, $69.2 \pm 7.6$, gives a prediction of
\wlng\ contribution in the signal region that is in agreement with the simulation-based
estimate. As with the \zllg\ estimate, the dominant uncertainty is theoretical.

The SM \ttg, VV\Pgg\ , \zllg, $\PW\to\ell\PGn$, and \gj\ processes are minor (${\sim} 10$\%)
backgrounds in the signal region. Although \zllg\ and \gj\ do not involve high-\PT invisible
particles, the former can exhibit large \met when the leptons are not reconstructed, and the latter
when jet energy is severely mismeasured. The estimates for all five processes are taken from
\MGfiveAMC simulations at leading order in QCD.

\subsection{Background estimation using recorded data}

An important background consists of $\PW\to\Pe\PGn$ events in which the electron is
misidentified as a photon. The misidentification occurs because of an inefficiency in seeding electron
tracks. A seeding efficiency of $\epsilon = 0.977 \pm 0.002$ for electrons with
$\PT > 160\GeV$ is measured in data using a tag-and-probe technique in $\PZ\to\Pe\Pe$ events, and is
verified with MC simulation. Misidentified electron events are modeled by a proxy sample of
electron events, defined in data by requiring an ECAL cluster with a pixel seed.
The proxy events must otherwise pass the same criteria used to select signal candidate events. The number of
electron proxy events is then scaled by $(1-\epsilon)/\epsilon$ to yield an estimated contribution
of $52.7 \pm 4.2$ events from electron misidentification. The dominant uncertainty in this estimate
is the statistical uncertainty in the measurement of $\epsilon$.

Electromagnetic showers from hadronic activity can also mimic a photon signature. This process is
estimated by counting the numbers of events in two different subsets of a low-\ptmiss multijet data
sample. The first subset consists of events with a photon candidate that satisfies the signal selection
criteria. These events contain both true photons and jets that are misidentified as photons. The
second subset comprises events with a candidate photon that meets less stringent shower-shape
requirements and inverted isolation criteria with respect to the signal candidates. Nearly all of
the candidate photons in these events arise from jet misidentification. The hadron misidentification
ratio is defined as the ratio between the number of the misidentified events in the first subset to
the total number of events in the second subset. The numerator is estimated by fitting the shower
shape distribution of the photon candidate in the first subset with template distributions. For true
photons, a template for the shower width is formed using simulated $\Pgg + \text{jets}$ events.  For
jets misidentified as photons, the template is obtained from a sample selected by inverting the
charged-hadron isolation and removing the shower-shape requirement entirely. Once the hadron
misidentification ratio is computed, it is multiplied by the number of events in the high-\ptmiss
control sample with a photon candidate that satisfies the conditions used to select the second subset
of the low-\ptmiss control sample. The product, $5.9 \pm 1.7$ events, is the estimate of the
contribution of jet misidentification background in the signal region. The dominant uncertanty is
systematic, and accounts for the effects of the fitting procedure, sample purity, photon candidate
definition of the control samples, and the sample bias in the jet composition.

Finally, backgrounds from beam halo and spikes in the ECAL are estimated from fits of the
angular and timing distributions of the calorimeter clusters. Energy clusters in the ECAL due to beam halo muons are observed to
concentrate around $\phi \sim 0\,\text{and}\,\pi$, while all other processes (collision-related
processes and ECAL spikes) produce photon candidates that are uniformly distributed in $\phi$. The
distribution of the cluster seed time provides a cross-check on this background estimate and an
independent means to estimate the ECAL spikes contribution. Exploiting these features, a
two-component fit of the $\phi$ distribution with beam halo and uniform templates, and a
three-component fit of the cluster seed time using the halo, spike, and prompt-photon templates are
performed. In both fits, the halo template is obtained by requiring high halo total energy for
candidate-like photon candidates. The timing distribution of the spike background is obtained by
inverting the shower shape requirement in the candidate photon selection. The results of the two fits are
combined into an uncertainty-weighted average. Beam halo and spike backgrounds of $5.5
\substack{+9.3\\-5.5}$ and $8.5 \pm 6.7$ events, respectively, are predicted, where the dominant
uncertainty is statistical.

\section{Results and interpretation}
\label{sec:results}

The estimated number of events and the associated uncertainty for each background process are given in Table~\ref{tab:BkgSummaryC}.
A total of 400
events are observed in data, which is in agreement with the total
expected SM background of $386 \pm 36$ events.

\begin{figure}[hbtp]
\centering
\includegraphics[width=0.495\linewidth]{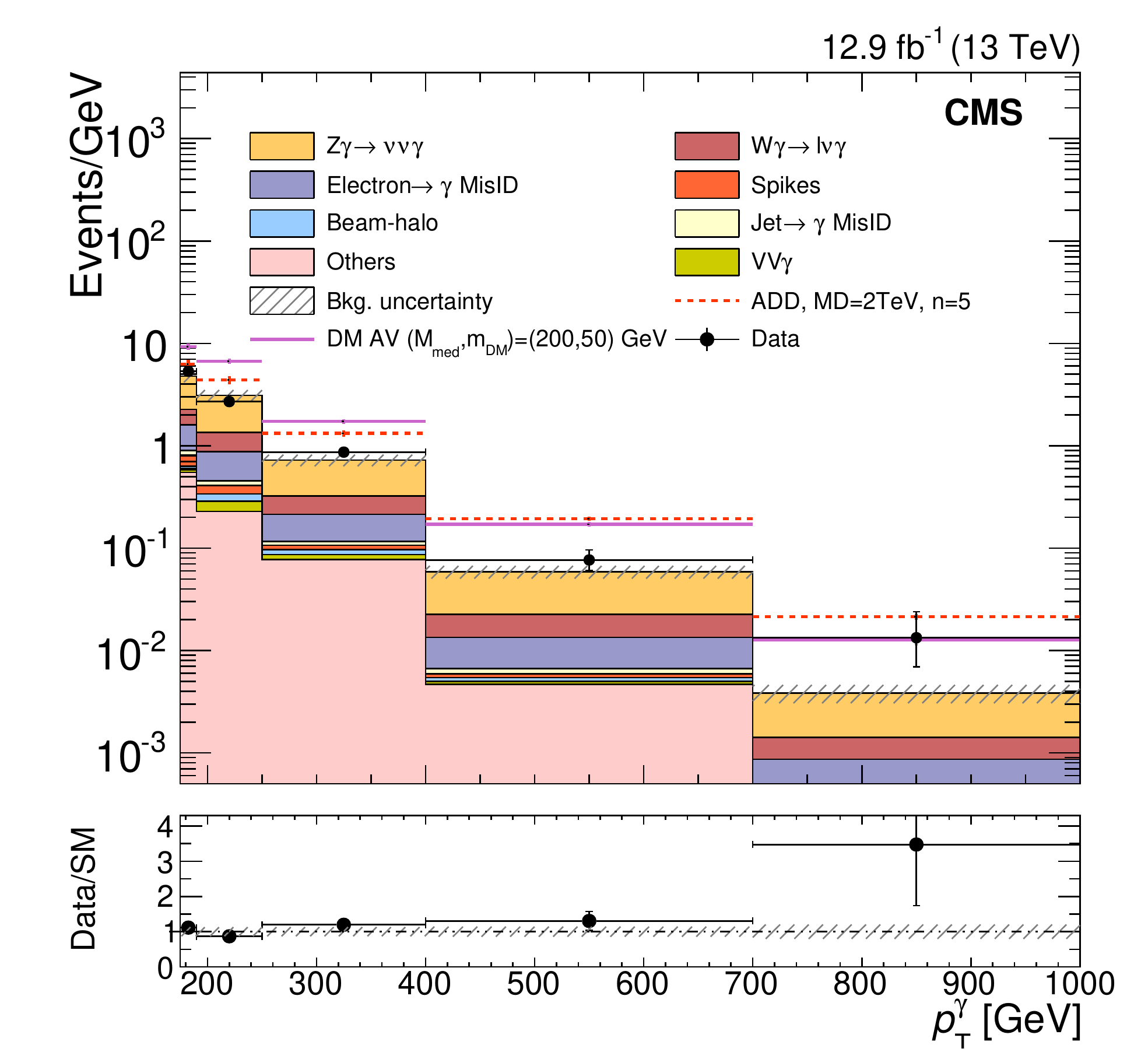}
\includegraphics[width=0.495\linewidth]{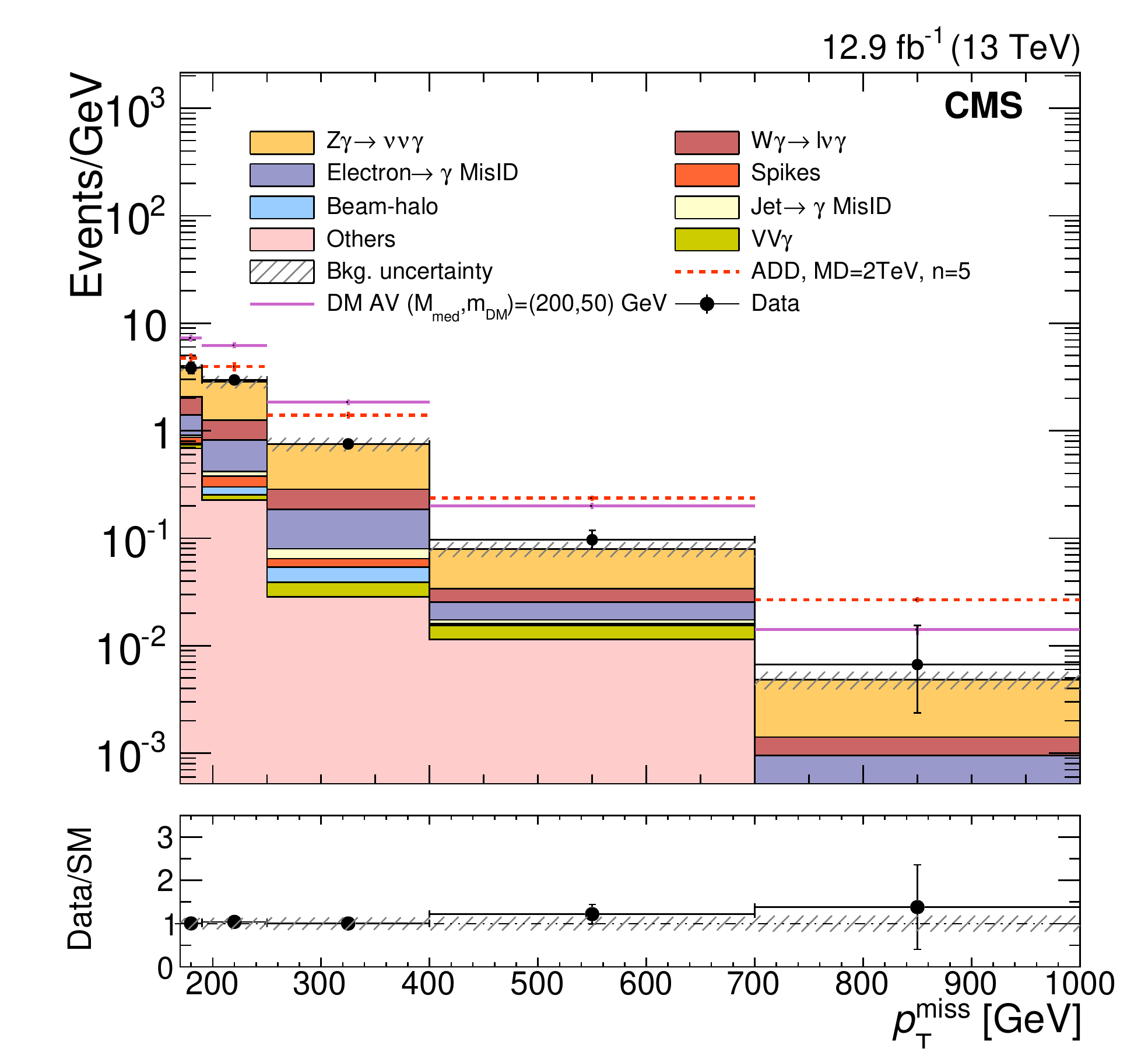}
\caption{
The \ETg\ (left) and \ptmiss\ (right) distributions for the candidate sample, compared with estimated contributions from SM backgrounds. In the legends, ``others'' includes the contribution from \gj, $\PW\to\ell\PGn$, \zllg, and \ttg backgrounds. The background uncertainties include statistical and systematic components. The last bin includes the overflow. The lower panel shows the ratio of data and SM background predictions, where the hatched band shows the systematic uncertainty.
}
\label{fig:stack_plot}

\end{figure}

Distributions of \ETg\ and \ptmiss for the
selected candidate events are shown in Fig.~\ref{fig:stack_plot} together with their respective estimated
background distributions. A summary of the systematic uncertainties for the background estimates is given in
Table~\ref{tab:sysfull1}. The quoted systematic uncertainties in Table~\ref{tab:sysfull1} follow the signal and background modeling
discussion in Section~\ref{sec:bkgestimate}.

\begin{table}[htbp]
\centering
\topcaption{
Summary of estimated background and observed candidate events. The quoted uncertainties for the background estimates are obtained by adding the systematic and statistical uncertainties in quadrature.
}
\label{tab:BkgSummaryC}
\begin{tabular}{l|c}
\hline
Process & Events \\
\hline
\zinvg & $215 \pm 32\x$  \\
\wlng & $57.2 \pm 8.0\x$    \\
Electron misidentification & $52.7 \pm 4.2\x$   \\
ECAL spikes &  $8.5 \pm 6.7$ \\
Beam halo &  $5.5\, \substack{+9.3\\-5.5}\x$ \\
$\Pgg + \text{jets}$ & $10.1 \pm 5.7\x$ \\
$\PW\to\Pgm\PGn$ & $8.5 \pm 3.0$  \\
\ttg & $8.2 \pm 0.6$  \\
Jet misidentification & $5.9 \pm 1.7$ \\
VV\Pgg & $5.5 \pm 1.8 $\\
$\PW\to\Pgt\PGn$ & $5.2 \pm 2.3$  \\
\zllg & $2.9 \pm 0.2$  \\
\hline
Total background & $386 \pm 36\x$ \\
\hline
Data & 400   \\
\hline
\end{tabular}

\end{table}

\begin{table}[htbp]
\centering
\topcaption{
Summary of relative systematic uncertainties (\%) for different background estimates. The middle column indicates the component of the estimated SM background that is affected by each uncertainty.
}
\label{tab:sysfull1}
\begin{tabular}{l|l|c}
\hline
Source & Background component & Value \\
\hline
Integrated luminosity~\cite{CMS-PAS-LUM-16-001} & All simulation-based & 6.2 \\
Jet and \Pgg\ energy scale, \met resolution & All simulation-based & 3--4 \\
Data/simulation factor & All simulation-based & 6 \\
PDF, \muR\ and \muF & \zinvg, \wlng & 5--9 \\
Electroweak higher-order corrections & \zinvg, \wlng & \x7--11 \\
Hadronic misidentification ratio & Jet misid. & 29 \\
Electron seeding $\epsilon$ & Electron misid. & 6 \\
ECAL spikes template shape & ECAL spikes & 75 \\
Beam halo template shape & Beam halo & $+169$/$-100$ \\
\gj\ yield & \gj & 54 \\
\hline
\end{tabular}

\end{table}

No excess of data with respect to the SM prediction is observed and limits are set on the
aforementioned DM and ADD models. The evaluation of systematic uncertainties for the simulated signal
follows the same procedures used for simulated backgrounds (Section~\ref{sec:bkgestimate}). For each
signal model, a 95\% confidence level (CL) cross section upper bound is obtained utilizing the
asymptotic CL$_\mathrm{s}$
criterion~\cite{Junk:1999kv,Read:2002hq,CMS-NOTE-2011-005,Cowan:2010js}. In this method, a Poisson
likelihood for the observed number of events is maximized under different signal strength
hypotheses, taking the systematic uncertainties as nuisance parameters that modify the signal and
background predictions. Each nuisance parameter is assigned a log-normal probability distribution,
using the systematic uncertainty value as the width. The best fit background predictions differ from
the original by at most 4\%. Confidence intervals are drawn by comparing these maximum likelihood values to
those computed from background-only and signal-plus-background pseudo-data.

\subsection{Limits on simplified dark matter models}

The simplified DM models proposed by the LHC Dark Matter Forum~\cite{dmforum} are designed to
facilitate the comparison and translation of various DM search results. In the models considered in
this analysis, Dirac DM particles couple to a vector or axial-vector mediator, which in turn couples
to the SM quarks. Model points are identified by a set of four parameters: the DM mass \mdm, the
mediator mass \mmed, the universal mediator coupling to quarks \gq, and the mediator coupling
to DM particles \gDM. In this analysis, we fix the values of \gq\ and
\gDM\ to 0.25 and 1.0, respectively, and scan the \mmed--\mdm\
plane~\cite{Boveia:2016mrp}. The search is not yet sensitive to the spin-0 mediator models defined
in Ref.~\cite{dmforum}.

\begin{figure}[htbp]
\centering
\includegraphics[width=0.48\linewidth]{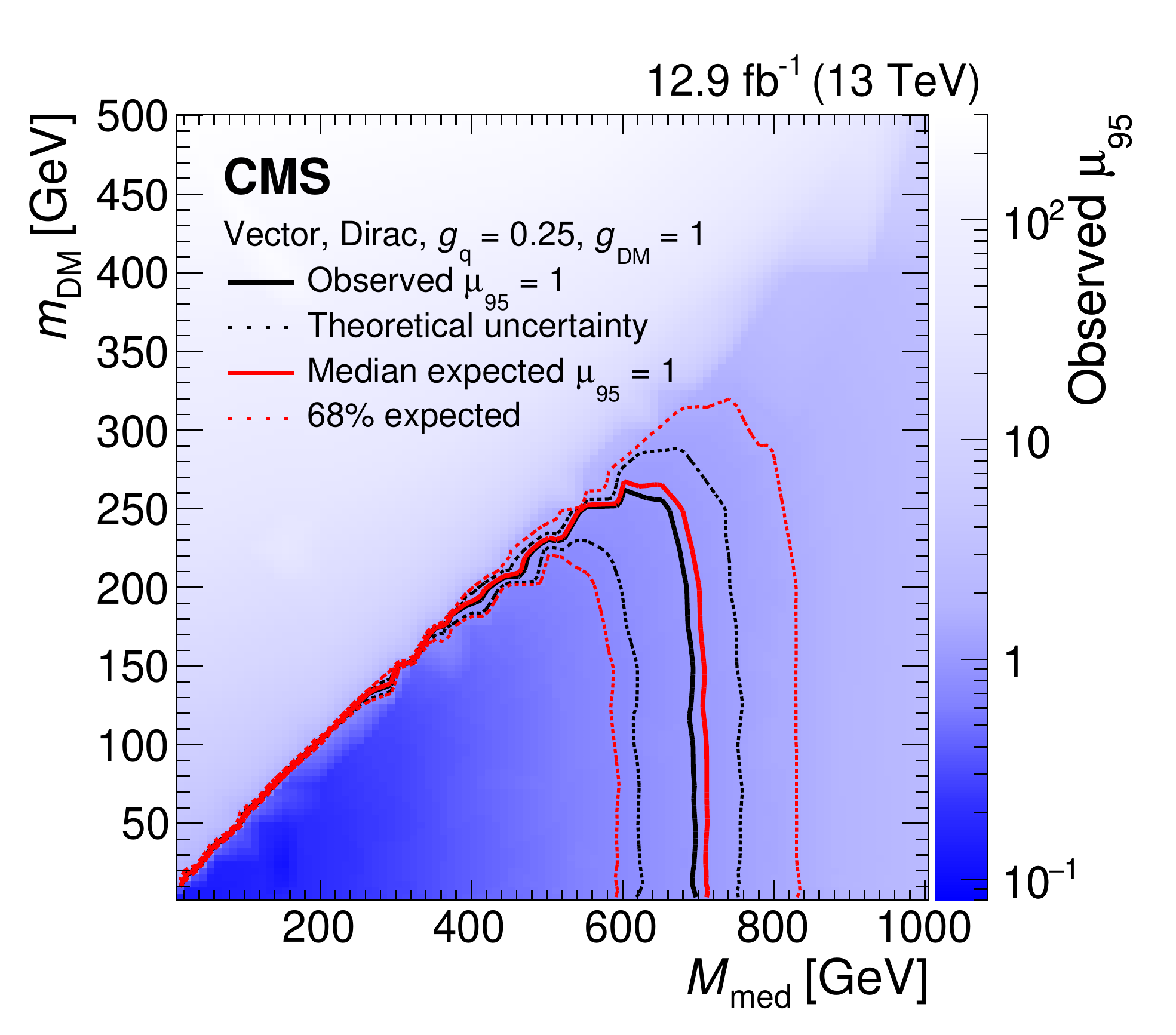}\quad
\includegraphics[width=0.48\linewidth]{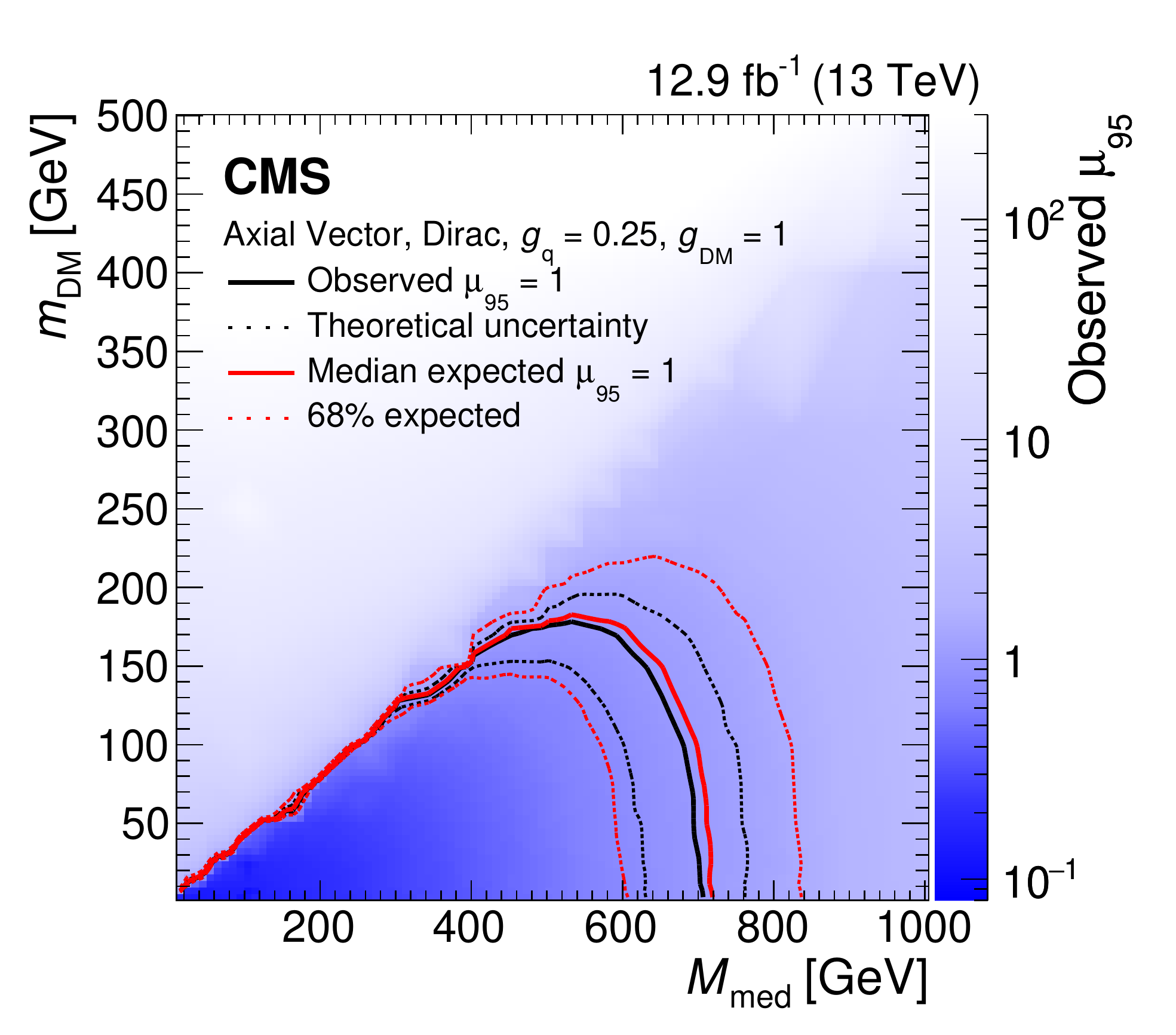}
\caption{
The ratio of 95\% CL cross section upper limits to theoretical cross section ($\mu_{95}$), for DM simplified models with vector (left) and axial-vector (right) mediators, assuming $\gq=0.25$ and $\gDM=1$.
Expected and observed $\mu_{95} = 1$ contours are overlaid. The region below the observed contour is excluded.
}
\label{fig:2d}

\end{figure}

Figure~\ref{fig:2d} shows the 95\% CL cross section upper limits with respect to the corresponding
theoretical cross section ($\mu_{95}= \sigma_{95\%}/\sigma_{\text{theory}}$) for the vector and
axial-vector mediator scenarios, in the \mmed--\mdm\ plane. The solid red (lighter) and black (darker) curves are the
expected and observed contours of $\mu_{95} = 1$ (exclusion contour). The region with $\mu_{95} < 1$ is excluded
under nominal $\sigma_{\text{theory}}$ hypotheses. The uncertainty in the expected upper limit includes the
experimental uncertainties. The uncertainty in the theoretical cross section is translated to the
uncertainty in the observed exclusion contour. While there is little difference in kinematic
properties between the two scenarios, the production cross section for heavier dark matter in the
vector mediator scenario tends to be higher~\cite{dmforum}, and therefore the exclusion region
broader. For the simplified DM models considered, mediator masses of up to 700\GeV are excluded for
small \mdm\ values.

The exclusion contours in Fig.~\ref{fig:2d} are also translated into the
$\sigma_{\text{SI/SD}}$--\mdm\ plane, where $\sigma_{\text{SI/SD}}$ are the
spin-independent/dependent DM-nucleon scattering cross sections. The translation and presentation of the result
follows the prescription given in Ref.~\cite{Boveia:2016mrp}.
In particular, to enable a direct comparison with results
from direct detection experiments, these limits are calculated at 90\% CL~\cite{dmforum}.
When compared to the direct detection experiments, the limits obtained from this search provide stronger constraints for dark matter masses less than 2\GeV, assuming spin-independent scattering, or less than 200\GeV, for spin-dependent scattering.

\begin{figure}[htbp]
\centering
\includegraphics[width=0.48\linewidth]{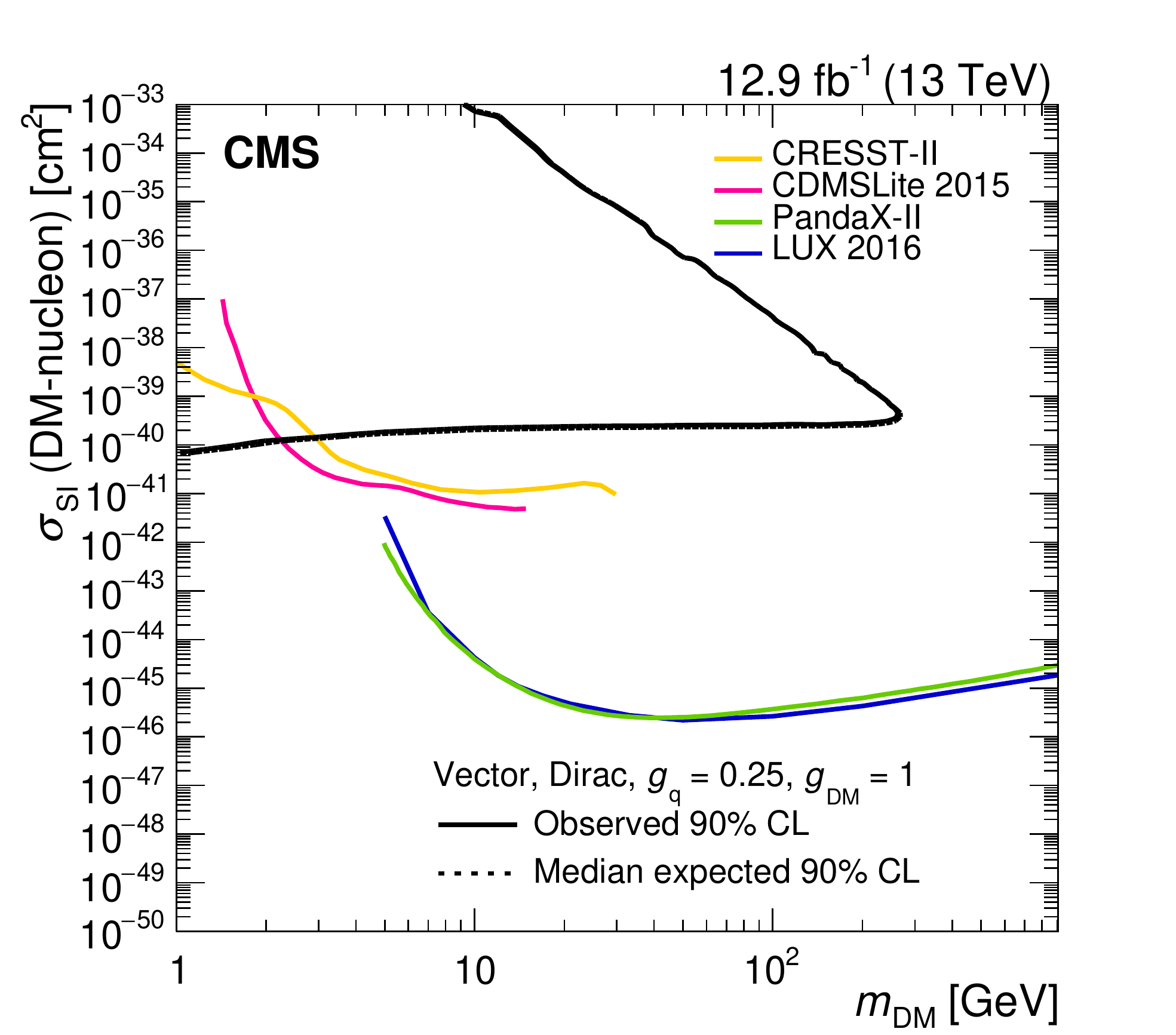}
\includegraphics[width=0.48\linewidth]{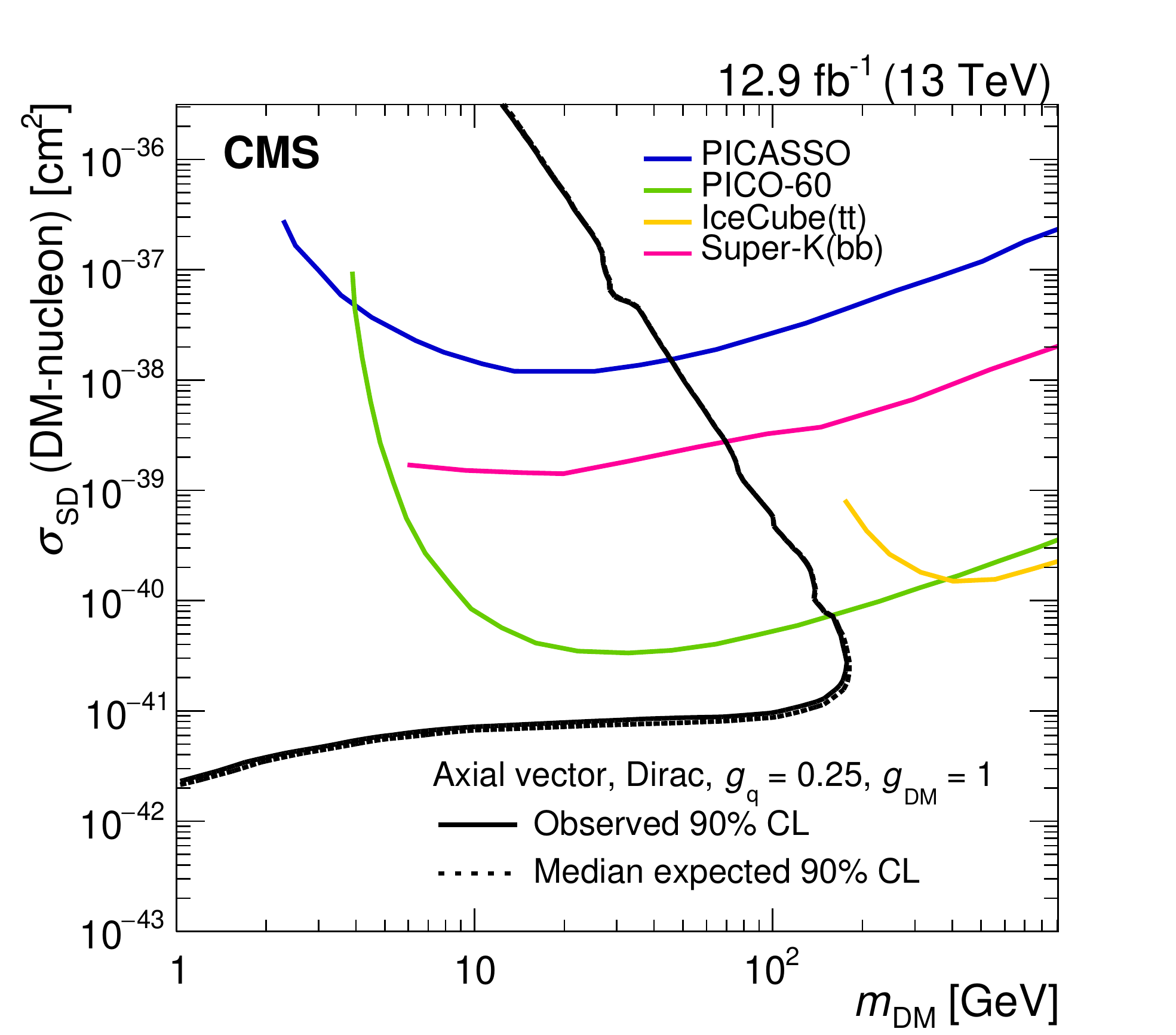}
\caption{
The 90\% CL exclusion limits on the $\chi$-nucleon spin-independent (left)
and spin-dependent (right) scattering cross sections involving vector and axial-vector operators, respectively, as a function of the \mdm.
Simplified model DM parameters of $\gq=0.25$ and $\gDM=1$ are assumed.
The region to the upper left of the contour is excluded. On the plots, the median expected 90\% CL curve overlaps the observed 90\% CL curve.
Also shown are
corresponding exclusion contours, where regions above the curves are excluded,
from the recent results by CDMSLite~\cite{Agnese:2015nto}, LUX~\cite{Akerib:2016vxi},
PandaX~\cite{Tan:2016zwf}, CRESST-II~\cite{Angloher:2015ewa}, PICO-60~\cite{Amole:2017dex},
IceCube~\cite{Aartsen:2016exj}, PICASSO~\cite{Behnke:2016lsk} and Super-Kamiokande~\cite{Choi:2015ara} Collaborations.
}
\label{fig:cont}

\end{figure}

\subsection{Limits on electroweak dark matter models}

\begin{figure}[htbp]
\centering
\includegraphics[width=0.48\linewidth]{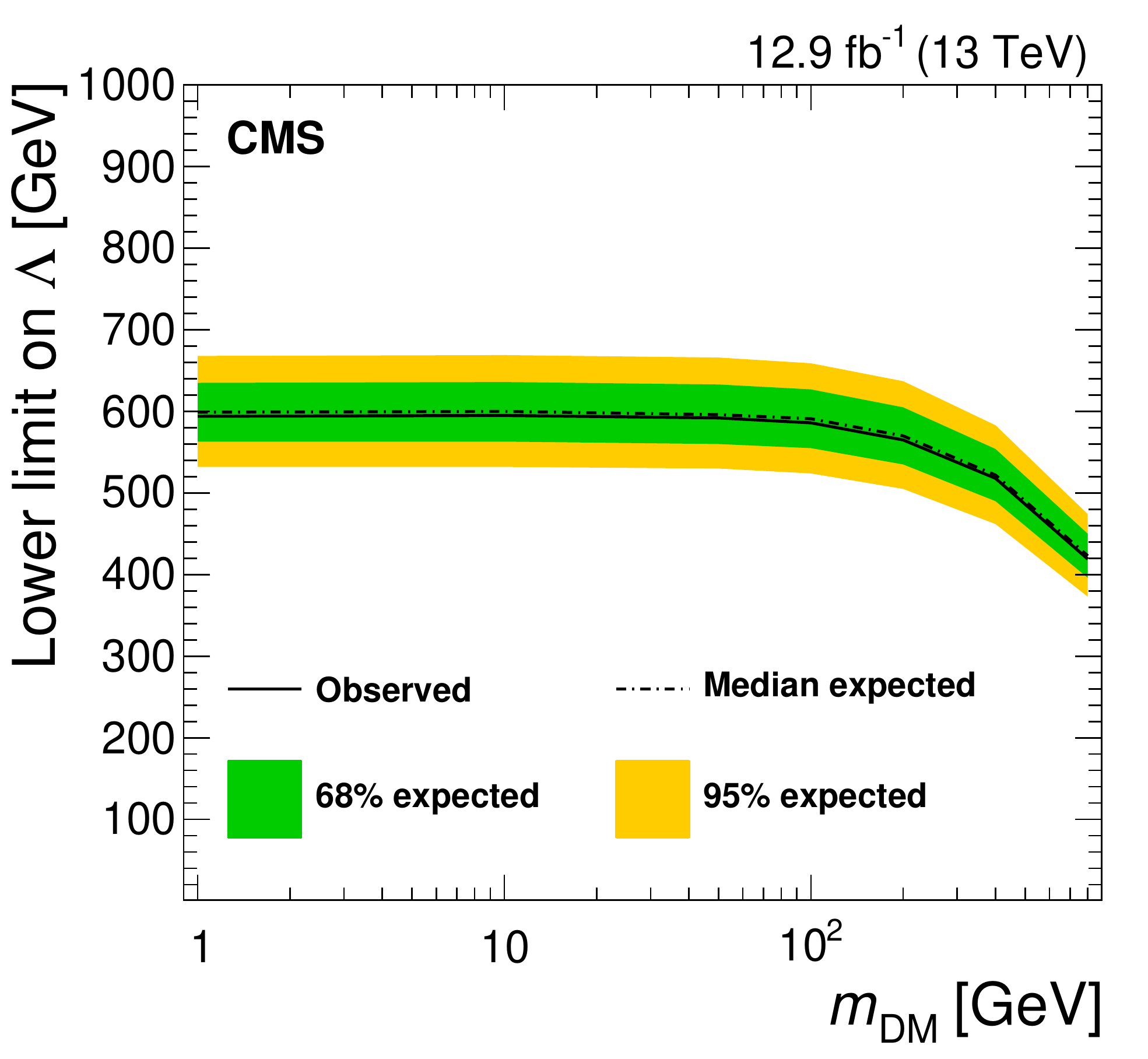}
\caption{
    The 95\% CL expected and observed lower limits on $\Lambda$ as a function of \mdm, for a dimension-7 operator EFT model assuming $k_{1} = k_{2} = 1$.
}
\label{fig:DMEWKlimits}

\end{figure}

The DM effective field theory (EFT) model contains a dimension-7 contact interaction of type
$\Pgg\Pgg\chi\overline{\chi}$~\cite{Nelson:2013pqa}. The interaction is described by four parameters: the
coupling to photons (parametrized in terms of coupling strengths $k_{1}$ and $k_{2}$), the DM
mass \mdm, and the suppression scale $\Lambda$. Since the interaction cross section is directly
proportional to $\Lambda^{-6}$, cross section upper limits are translated into lower limits on
$\Lambda$, assuming $k_{1} = k_{2} = 1$. The expected and observed lower limits on $\Lambda$ as a function of \mdm\ are shown in
Fig.~\ref{fig:DMEWKlimits}. Values of $\Lambda$ up to 600\GeV are excluded at 95\% CL.

\subsection{Limits on the ADD model}
\label{ssec:lim_ADD}

\begin{figure}[htbp]
\centering
\includegraphics[width=0.5\textwidth]{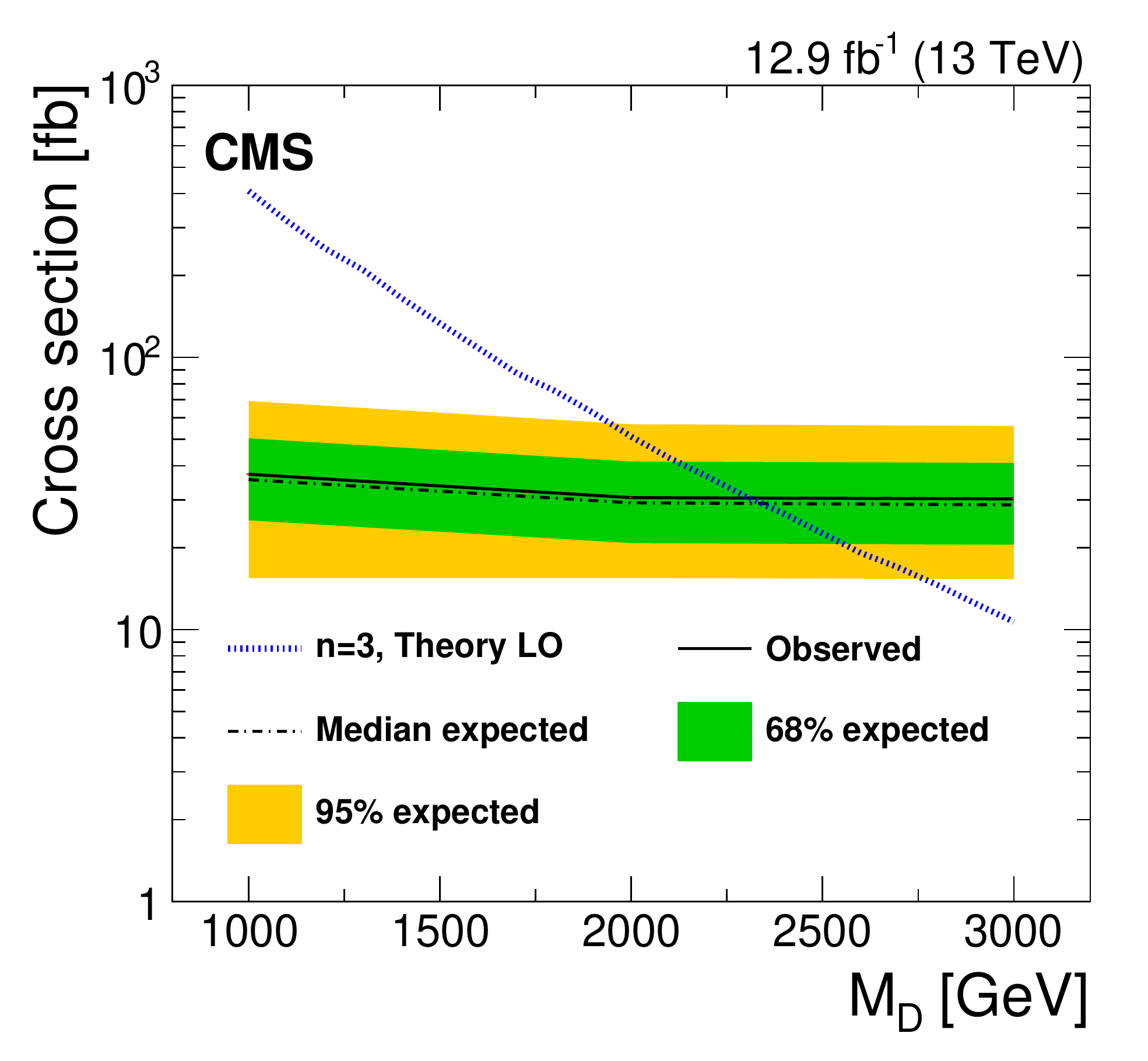}
\caption{The 95\% CL upper limits on the ADD graviton production cross section, as a function of \mD\, for $n=3$ extra dimensions.}
\label{fig:ADDLimits}

\end{figure}

\begin{figure}[htbp]
\centering
\includegraphics[width=0.5\textwidth]{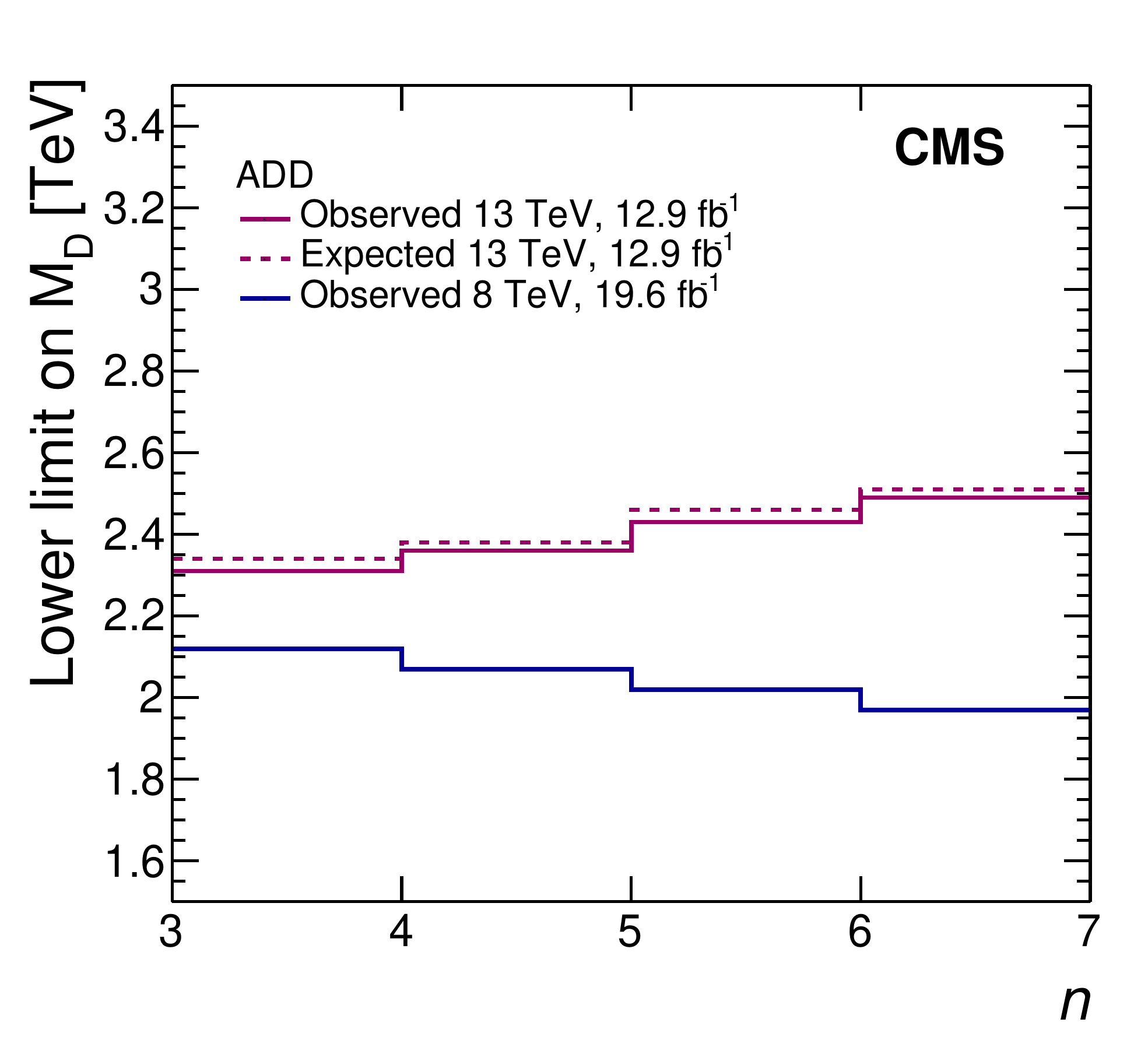}
\caption{Lower limit on \mD\ as a function of $n$, the number of ADD extra dimensions.}
\label{fig:MDLimits}

\end{figure}

Figure~\ref{fig:ADDLimits} shows the upper limit and the theoretically calculated ADD graviton
production cross section for $n=3$ extra dimensions, as a function of \mD. Lower limits on \mD\ for
various values of $n$ extra dimensions are summarized in Table~\ref{tab:MDLimits}, and in Fig.~\ref{fig:MDLimits} are compared
to CMS results at $\sqrt{s}=8\TeV$~\cite{mono8TeV}.
Because the graviton production cross section
scales as $E^n/\mD^{n+2}$~\cite{Giudice:1998ck}, where $E$ is the
typical energy of the hard scattering, \mD\ can be an increasing or
decreasing function of $n$ for a fixed cross section value,
approaching $E$ as $n\to \infty$. Note that the value of $E$
is dependent on the center-of-mass energy of the \Pp\Pp\ collision,
and is ${\sim} 2\TeV$ for $\sqrt{s}=8\TeV$ and ${\sim} 3\TeV$ for $\sqrt{s}=13\TeV$.
Values of \mD\ up to 2.49\TeV for $n=6$ are excluded by the current analysis.

\begin{table}[htbp]
\centering
\topcaption{The 95\% CL observed and expected lower limits on \mD\ as a function of $n$, the number of ADD extra dimensions.}
\label{tab:MDLimits}
\begin{tabular}{c|c|c}
\hline
$n$ & Obs. limit (\TeVns{}) & Exp. limit (\TeVns{}) \\
\hline
3 & 2.31 & 2.34 \\
4 & 2.36 & 2.38 \\
5 & 2.43 & 2.46\\
6 & 2.49 & 2.51 \\
\hline
\end{tabular}

\end{table}

\section{Summary}
\label{sec:summary}

Proton-proton collisions producing large missing transverse momentum and
a high transverse momentum photon have been investigated to search for
new phenomena, using a data set corresponding to \currentL\fbinv of
integrated luminosity recorded at $\sqrt{s} = 13\TeV$ at the CERN LHC. No deviations from the
standard model predictions are observed. Constraints are set on the
production cross sections for dark matter and large extra dimension
gravitons at 95\% confidence level, which are then translated to
limits on the parameters of the individual models. For the simplified
dark matter production models considered, the search excludes mediator
masses of up to 700\GeV for low-mass dark matter. For an effective
dimension-7 photon-dark matter contact interaction, values of
$\Lambda$ up to 600\GeV are excluded.  For the ADD model with extra
spatial dimensions, values of the fundamental Planck scale up to $2.31
\text{--} 2.49\TeV$, depending on the number of extra dimensions, are
excluded. These are the most stringent limits to date using the monophoton final state.

\begin{acknowledgments}

We congratulate our colleagues in the CERN accelerator departments for the excellent performance of the LHC and thank the technical and administrative staffs at CERN and at other CMS institutes for their contributions to the success of the CMS effort. In addition, we gratefully acknowledge the computing centres and personnel of the Worldwide LHC Computing Grid for delivering so effectively the computing infrastructure essential to our analyses. Finally, we acknowledge the enduring support for the construction and operation of the LHC and the CMS detector provided by the following funding agencies: BMWFW and FWF (Austria); FNRS and FWO (Belgium); CNPq, CAPES, FAPERJ, and FAPESP (Brazil); MES (Bulgaria); CERN; CAS, MoST, and NSFC (China); COLCIENCIAS (Colombia); MSES and CSF (Croatia); RPF (Cyprus); SENESCYT (Ecuador); MoER, ERC IUT, and ERDF (Estonia); Academy of Finland, MEC, and HIP (Finland); CEA and CNRS/IN2P3 (France); BMBF, DFG, and HGF (Germany); GSRT (Greece); OTKA and NIH (Hungary); DAE and DST (India); IPM (Iran); SFI (Ireland); INFN (Italy); MSIP and NRF (Republic of Korea); LAS (Lithuania); MOE and UM (Malaysia); BUAP, CINVESTAV, CONACYT, LNS, SEP, and UASLP-FAI (Mexico); MBIE (New Zealand); PAEC (Pakistan); MSHE and NSC (Poland); FCT (Portugal); JINR (Dubna); MON, RosAtom, RAS, RFBR and RAEP (Russia); MESTD (Serbia); SEIDI, CPAN, PCTI and FEDER (Spain); Swiss Funding Agencies (Switzerland); MST (Taipei); ThEPCenter, IPST, STAR, and NSTDA (Thailand); TUBITAK and TAEK (Turkey); NASU and SFFR (Ukraine); STFC (United Kingdom); DOE and NSF (USA).

\hyphenation{Rachada-pisek} Individuals have received support from the Marie-Curie programme and the European Research Council and Horizon 2020 Grant, contract No. 675440 (European Union); the Leventis Foundation; the A. P. Sloan Foundation; the Alexander von Humboldt Foundation; the Belgian Federal Science Policy Office; the Fonds pour la Formation \`a la Recherche dans l'Industrie et dans l'Agriculture (FRIA-Belgium); the Agentschap voor Innovatie door Wetenschap en Technologie (IWT-Belgium); the Ministry of Education, Youth and Sports (MEYS) of the Czech Republic; the Council of Science and Industrial Research, India; the HOMING PLUS programme of the Foundation for Polish Science, cofinanced from European Union, Regional Development Fund, the Mobility Plus programme of the Ministry of Science and Higher Education, the National Science Center (Poland), contracts Harmonia 2014/14/M/ST2/00428, Opus 2014/13/B/ST2/02543, 2014/15/B/ST2/03998, and 2015/19/B/ST2/02861, Sonata-bis 2012/07/E/ST2/01406; the National Priorities Research Program by Qatar National Research Fund; the Programa Clar\'in-COFUND del Principado de Asturias; the Thalis and Aristeia programmes cofinanced by EU-ESF and the Greek NSRF; the Rachadapisek Sompot Fund for Postdoctoral Fellowship, Chulalongkorn University and the Chulalongkorn Academic into Its 2nd Century Project Advancement Project (Thailand); and the Welch Foundation, contract C-1845.

\end{acknowledgments}

\bibliography{auto_generated}
\cleardoublepage \appendix\section{The CMS Collaboration \label{app:collab}}\begin{sloppypar}\hyphenpenalty=5000\widowpenalty=500\clubpenalty=5000\textbf{Yerevan Physics Institute,  Yerevan,  Armenia}\\*[0pt]
A.M.~Sirunyan, A.~Tumasyan
\vskip\cmsinstskip
\textbf{Institut f\"{u}r Hochenergiephysik,  Wien,  Austria}\\*[0pt]
W.~Adam, E.~Asilar, T.~Bergauer, J.~Brandstetter, E.~Brondolin, M.~Dragicevic, J.~Er\"{o}, M.~Flechl, M.~Friedl, R.~Fr\"{u}hwirth\cmsAuthorMark{1}, V.M.~Ghete, C.~Hartl, N.~H\"{o}rmann, J.~Hrubec, M.~Jeitler\cmsAuthorMark{1}, A.~K\"{o}nig, I.~Kr\"{a}tschmer, D.~Liko, T.~Matsushita, I.~Mikulec, D.~Rabady, N.~Rad, B.~Rahbaran, H.~Rohringer, J.~Schieck\cmsAuthorMark{1}, J.~Strauss, W.~Waltenberger, C.-E.~Wulz\cmsAuthorMark{1}
% ...
\vskip\cmsinstskip
% ...
1:~~Also at Vienna University of Technology, Vienna, Austria\\
% ...

\end{sloppypar}
\end{document}